\documentclass[]{spie}  

\usepackage[T1]{fontenc}
\usepackage{amsmath,amsfonts,amssymb}
\usepackage{upgreek}
\usepackage{graphicx}
\usepackage[colorlinks=true, allcolors=blue]{hyperref}

\usepackage{xcolor}

\newcommand{\GP}[1]{\left(#1\right)}

\newcommand{\upgrades}[1]{\textcolor{red}{#1}} 

\usepackage[inkscapeformat=png]{svg}

\title{Photonic spectro-interferometry with SCExAO/FIRST at the Subaru Telescope: towards H{\Large{$\boldsymbol{\alpha}$}} imaging of protoplanets}

\author[a,b]{S.~Vievard}
\author[a,c]{M.~Lallement}
\author[c]{E.~Huby}
\author[c]{S.~Lacour}
\author[a,b,d]{O.~Guyon}
\author[e]{N.~Jovanovic}
\author[f]{S.~Leon-Saval}
\author[a]{J.~Lozi}
\author[a]{V.~Deo}
\author[a]{K.~Ahn}
\author[g]{N.~Cvetojevic}
\author[c]{K.~Barjot}
\author[h]{G.~Martin}
\author[c]{H.D. Kenchington-Goldsmith}
\author[h,o]{G.~Duchêne}
\author[b,n]{T.~Kotani}
\author[p]{F.~Marchis}
\author[c]{D.~Rouan}
\author[l]{M.~Fitzgerald}
\author[i]{S.~Sallum}
\author[j,k]{B.~Norris}
\author[f]{C.~Betters}
\author[e]{P.~Gatkine}
\author[l]{J.~Lin}
\author[l]{Y.J.~Kim}
\author[m]{C. Pham}
\author[m]{C. Cassagnettes}
\author[m]{A. Billat}
\author[b,q,r]{M. Tamura}
\author[c]{G.~Perrin}

\affil[a]{\small Subaru Telescope, National Astronomical Observatory of Japan, National Institutes of Natural Sciences (NINS), 650 North A'oh\=ok\=u Place, Hilo, HI, 96720, U.S.A.}
\affil[b]{Astrobiology Center of NINS, 2-21-1, Osawa, Mitaka, Tokyo, 181-8588, Japan}
\affil[c]{LESIA, Observatoire de Paris, Universit\'e PSL, CNRS, Sorbonne Universit\'e, Sorbonne Paris Cite\'e, 5 place Jules Janssen, 92195 Meudon, France }
\affil[d]{College of Optical Sciences, University of Arizona, Tucson, AZ 85721, U.S.A.}
\affil[e]{California Institute of Technology, 1200 E California Blvd, Pasadena, CA 91125, U.S.A.}
\affil[f]{Sydney Astrophotonic Instrumentation Laboratory, The University of Sydney, Sydney, NSW 2006, Australia}
\affil[g]{Observatoire de la C\^ote d'Azur, 96 Boulevard de l'Observatoire, 06300 Nice, France}
\affil[h]{Univ. Grenoble Alpes, CNRS, IPAG, 38000 Grenoble, France}
\affil[i]{Univ. of California, Irvine, G302 C Student Center, Irvine, CA 92697}
\affil[j]{Sydney Institute for Astronomy, School of Physics, The University of Sydney, NSW 2006, Australia}
\affil[k]{AAO-USyd, School of Physics, University of Sydney 2006}
\affil[l]{Univ. of California, Los Angeles, 405 Hilgard Avenue, CA 90095}
\affil[m]{TEEM Photonics, 61 Chem. du Vieux Chêne, Meylan, France}
\affil[n]{Department of Astronomical Science, The Graduate Un iversity for Advanced Studies, SOKENDAI, 2-21-1 Osawa, Mitaka, Tokyo 181-8588, Japan}
\affil[o]{Department of Astronomy, University of California at Berkeley, Berkeley, CA 94720-3411}
\affil[p]{Carl Sagan Center at the SETI Institute, 189 Bernardo Av., Mountain View, CA 94043, USA}
\affil[q]{National Astronomical Observatory of Japan, 2-21-1 Osawa, Mitaka, Tokyo 181-8588, Japan}
\affil[r]{Department of Astronomy, Graduate School of Science, The University of Tokyo, Tokyo, Japan}

\authorinfo{Further author information: Sebastien Vievard: E-mail: vievard@naoj.org}

\begin{document} 
\maketitle

\begin{abstract}
FIRST is a post Extreme Adaptive-Optics (ExAO) spectro-interferometer operating in the Visible (600-800 nm, R$\sim$400). Its exquisite angular resolution (a sensitivity analysis of on-sky data shows that bright companions can be detected down to 0.25$\lambda$/D) combined with its sensitivity to pupil phase discontinuities (from a few nm up to dozens of microns) makes FIRST an ideal self-calibrated solution for enabling exoplanet detection and characterization in the future. We present the latest on-sky results along with recent upgrades, including the integration and on-sky test of a new spectrograph (R$\sim$3,600) optimized for the detection of H$\alpha$ emission from young exoplanets accreting matter.

\end{abstract}


\section{INTRODUCTION}
\label{sec:intro}  

Mass accretion signatures from protoplanets are the best avenue to study how planets form in protoplanetary disks. The mass accretion rate is a fundamental unknown property, which is currently best estimated by observations of hydrogen lines such as H$\alpha$. Two different models~\cite{aoyama2018theoretical,aoyama2021comparison} can explain the mechanism of such emission lines : 1- the shock model, where the hydrogen is emitted by the shock of the gas accreting onto the planet surface, and 2- the accretion-flow model, where the hydrogen is emitted by the hot accretion flow. The two models lead to large differences in the estimation of planetary accretion rate. While very high resolution spectroscopy (<10km/s or R>30,000) is required to discriminate between these line emission mechanisms, only a couple thousands resolution is enough for the H$\alpha$ line detection. To this day, only three accreting planets embedded in their disk were published, with contrasts down to 1:1000~\cite{haffert2019two,currie2022images}. It is required to identify more accreting planets for statistical discussions, by using high contrast imaging. 


The detection of protoplanets requires both high angular resolution and high contrast capabilities. Extreme Adaptive Optics (ExAO) systems coupled with coronagraphs already demonstrated their ability to fulfill both requirements, with an inner working angle of a few resolution elements. Such instruments on 10-m class telescopes are unfortunately unable to probe planets within 0.1~arcsecond (or 10~AU) from the host star. A way to explore these circumstellar regions is to use single aperture interferometry. The interferometric recombination of multiple subpupils from a single telescope can perform high contrast imaging with an inner working angle down to at least half the theoretical angular resolution of the telescope. Moreover, spectro-interferometers are by design sensitive to phasing errors in the telescope pupil, in particular in the presence of the Low Wind Effect (LWE), where large discontinuous aberrations are induced by radiative exchanges between the telescope spider and ambient air. The latter are invisible to the main ExAO control loop, and degrade the performance of current direct imaging techniques using coronagraphy.

The Subaru Coronagraphic Extreme Adaptive Optics\cite{2015PASP..127..890J} (SCExAO) platform hosts a single aperture interferometer, coupled with a spectrograph: FIRST\cite{perrin2006high,huby2012first,vievard2023singleaperture} (Fibered Imager foR a Single Telescope), operating in the Visible (650-780~nm, R$\sim$400), and based on pupil remapping using single mode fibers (SMFs). FIRST is routinely used on-sky and is subject to continuous upgrades. In this paper we first present the principle of the instrument, its integration at the Subaru Telescope, and its main scientific achievements. Second, we present various on-going upgrades aiming at increasing the instrument capabilities in order to identify accreting planet via H$\alpha$ imaging.

\section{Fibered spectro-interferometry at the Subaru Telescope}

\subsection{Principle}
\begin{figure}[!b]
	\centering
	\includegraphics[width=0.9\linewidth]{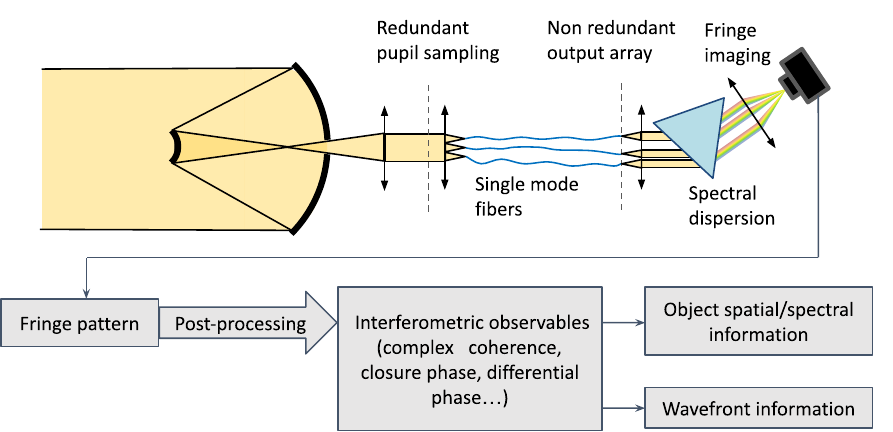}\\
	\caption{FIRST simplified principle. The pupil light is sampled thanks to a micro-lens array. The light from each sample, or subpupil is injected into a single-mode fiber. The single-mode fibers are rearranged into a 1D non-redundant array. Outputs of the fibers are dispersed and interfere on the science detector. Interferometric observables are extracted from the fringe pattern, allowing to obtain object's spatial/spectral information along with wavefront information.}
	\label{fig-firstpp}
\end{figure}

FIRST aims at using a single telescope as a coherent interfer-
ometer. The instrument concept is shown in Fig.~\ref{fig-firstpp}.

{}

The telescope pupil is sampled using a micro-lens array. Each sample is injected into a SMF which filters high order perturbations of the wavefront thanks to their mono-mode nature. The SMFs are then re-arranged into a non-redundant configuration, allowing to retrieve each baseline signal independently in their resulting interferometric recombination. A prism disperses the interferometric signal, providing spectral information. Using the fringe fitting technique\cite{tatulli2007interferometric}, we estimate each baseline complex coherence to extract either the astronomical object's spatial/spectral information or wavefront information. 

\subsection{FIRST on SCExAO}

Integrated on SCExAO, FIRST operates in the Visible wavelength range from about 600~nm to 800~nm. FIRST is composed of two separate modules: the injection and the recombination. \\


\subsubsection{The injection module}
\begin{figure}[!h]
    \centering
    \includegraphics[width=0.85\linewidth]{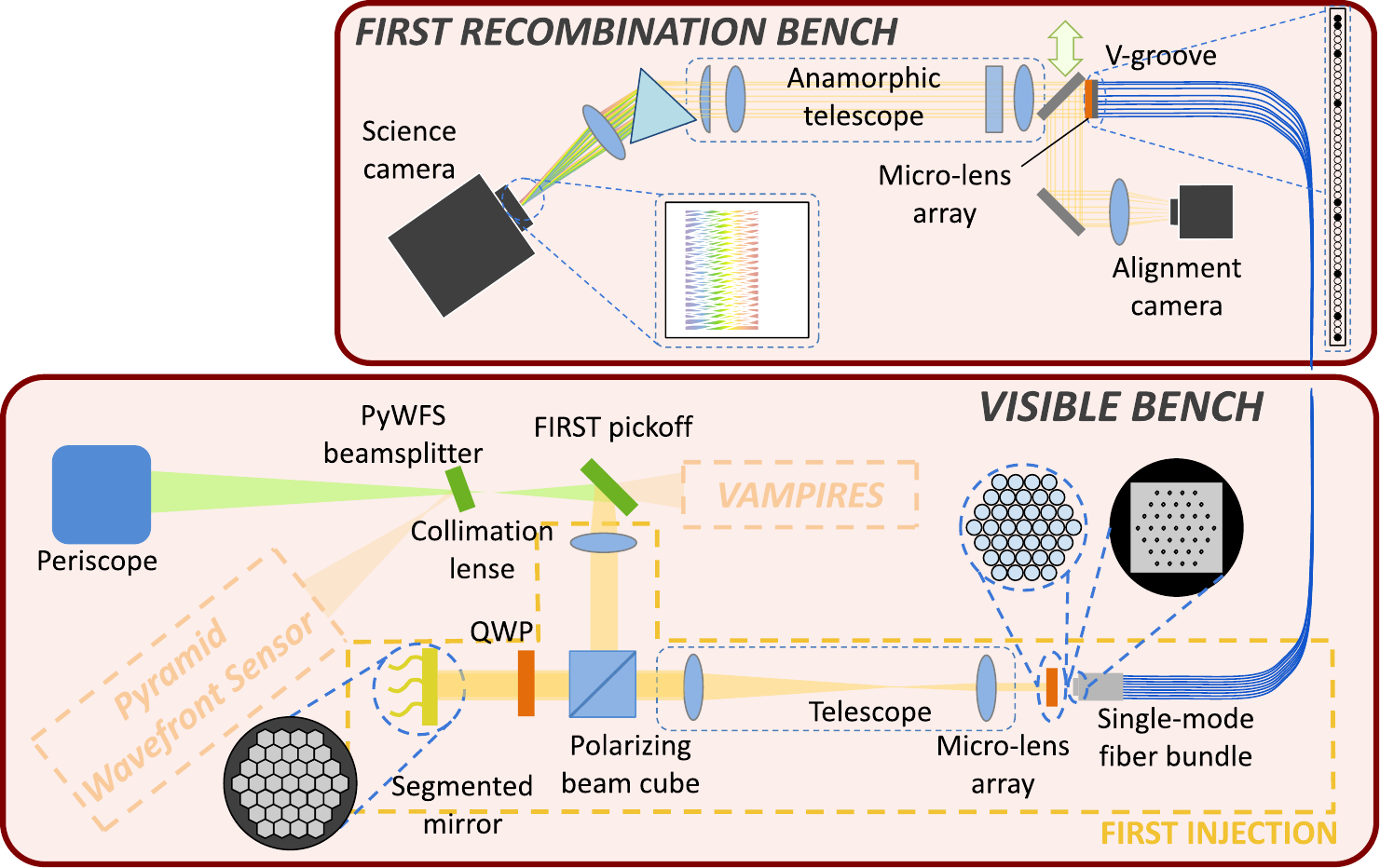}
    \caption{FIRST injection bench on the SCEXAO Visible bench (bottom), and FIRST recombination bench (top).}
    \label{fig:first-scexao-scheme}
\end{figure}

The injection module is located on the visible bench of SCExAO. We show the optical path of the instrument in Figure~\ref{fig:first-scexao-scheme}. A micro-lens array (MLA) is used to sample the pupil into 37 subpupils. A bundle of 37 single-mode fibers is placed in the focal plane of the MLA. FIRST exploits 9 of these subpupils, and their injection into the SMFs is optimized thanks to a 37-segmented mirror (Micro-ElectroMechanical System technology or MEMS - vendor: Iris AO). These 9~fibers are connected to manually polished SMF extensions to ensure matching path lengths. The extension fibers run from the injection bench to the recombination bench. 

\subsubsection{The recombination module}
\label{sec-firstrecomb}
The recombination module sits on its own bench, separated from SCExAO. Within the recombination module, the fibers outputs are connected to a linear V-groove, in a non-redundant configuration. Another MLA collimates the outputs of the V-groove fibers. We then have two paths for the light: a photometric monitoring path and a science path. In the photometric monitoring path, the output of the V-groove is imaged on a camera. An automatic procedure adjusts the MEMS segment tip/tilts to maximize the flux. Once the injection is optimized, the beams from the V-groove output can be directed to the science path. In the latter, the beams are spectrally dispersed by a combination of anamorphic optics, shaping the beams to mimic the spectrograph slit, and a prism, before being recombined on the science camera. 

{}

\subsection{Science with FIRST}

\begin{figure}[!h]
    \centering
    \includegraphics[width=0.85\linewidth]{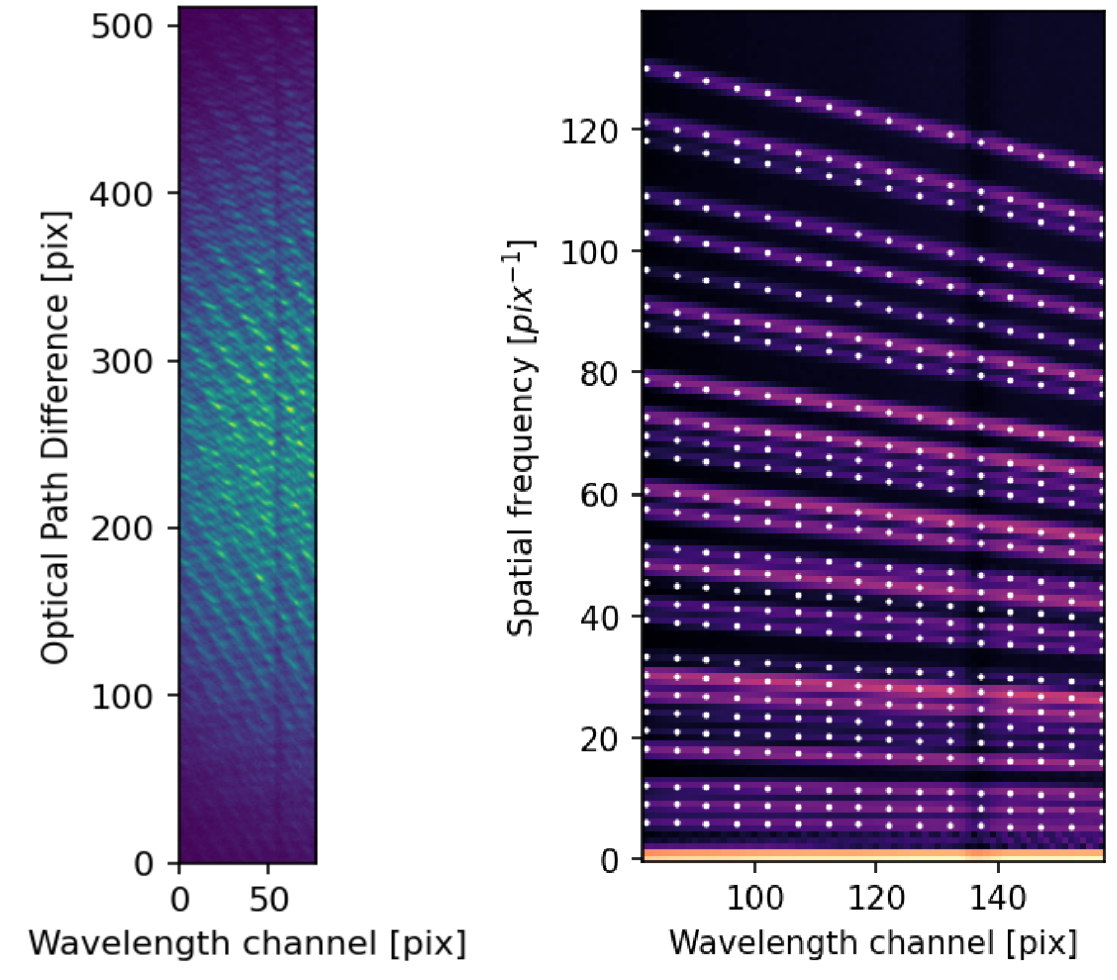}
    \caption{Typical image delivered by FIRST. Left: Exposure of 100 ms and 50 EMgain of Keho`oea, corrected from curvature. Right: Power spectral density computed from 30 cubes of 100 acquisitions with 100 ms exposure and 50 EMgain. The white dots show the PSD peak fitting for some selected wavelengths.}
    \label{Vega_imDPS}
\end{figure}

An example of an interferogram we obtain with FIRST is shown in Fig.~\ref{Vega_imDPS}-left. The spectral dispersion is here horizontal and the optical path difference is vertical. This image of 100 ms exposure and EM gain of 50 was obtained on Keho`oea [Vega]. Figure~\ref{Vega_imDPS}-right shows the power spectral density where we can see the spatial frequencies for each spectral channel. The non-redundant configuration of the fiber outputs allows for each spatial frequency to be isolated from the others. The data reduction pipeline, allowing to extract the interferometric observables from the fringes, has been previously derived~\cite{huby2012first} and won't be detailed here. 

{}

\subsubsection{Unresolved star observation}
We observed Keho`oea on July 13th 2020 UTC. We obtained a total of 5000 images with an integration time of 100~ms and an EM gain of 50 (an example of such image is shown in Figure~\ref{Vega_imDPS}). We extracted the complex coherence and closure phase measurements from the data. After calibration of the known instrumental bias, the closure phase values are (close to) zero - as expected for an unresolved star observation. Detailed data analysis and results are presented in Vievard et al. (2023)~\cite{vievard2023singleaperture} .

\begin{figure}[!h]
    \centering
    \includegraphics[width=\linewidth]{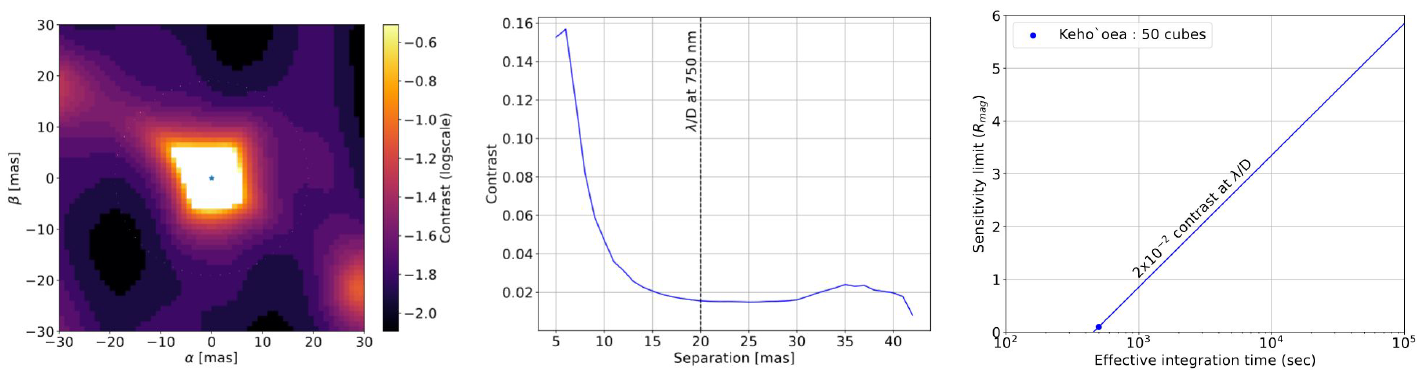}
    \caption{Performance of the FIRST instrument, estimated from unresolved source observation. Left : $3\sigma$ detection map; Middle: Computation of the left map radial profile (excluding 0-values); Right: Extrapolation of the detection performance for different star
magnitude as a function of the effective integration time.}
    \label{fig:vegaTest}
\end{figure}

We used these data to estimate FIRST performance in terms of achievable contrast and spatial resolution. To do so, we follow the methodology described in Absil et al. (2011)~\cite{absil2011searching} and Le Bouquin et al. (2012)\cite{le2012sensitivity} where interferometric data of an unresolved source are converted into a probability of companionship detection around the source considered. We then built a detection map around Keho`oea, where a $3\sigma$ detection would be possible. The result is shown in Figure~\ref{fig:vegaTest} (left and middle). From the detection map, we compute a contrast curve as a function of the separation to the central star. Results show first that a $3\sigma$ detection could be possible for a companion with a contrast down to $2\times 10^{-2}$ at the resolution power of the telescope. Furthermore, we see that companionship detection would be possible down to 5~mas around the star.

{}

Figure~\ref{fig:vegaTest}-right shows an extrapolation of the previous detection limit result to estimate the instrument performance for stars with
different magnitudes. It consists in estimating the effective integration time $\tau$, that would be required to obtain the same performance as on Keho`oea ($R_{mag,0} = 0.1$ observed with an integration time $\tau_0$). To do so, we use the following equation:
\begin{equation}\label{eq-Rmaglim}
    R_{mag} = R_{mag, 0} + 2.5\text{log}\GP{\frac{\tau}{\tau_0}},
\end{equation}
This extrapolation is valid under the assumptions of 1- similar observation conditions, and 2- photon noise limited regime. 
The graph, presenting the magnitude limit as a function of the effective integrated time, allows us to convey that FIRST can, at this stage, reach contrasts of about 0.02, around magnitude 3 stars, with an effective exposure time of about 2~hours.

\subsubsection{Binary star observations}
We observed Hokulei ($\alpha$ Aurigae, mR = -0.52, and semi-major axis = 56.4 mas) on September 16th and 17th 2020 UTC. We obtained 19 cubes of 500 frames, on each night, with an exposure time/EMgain of 10~ms/300 on the first night and 40~ms/300 on the second night. The detailed data analysis is presented in Vievard et al.(2023)~\cite{vievard2023singleaperture}. Using the closure phase fitting technique, we estimated the position of the Hokulei AB binary component for both epochs. The data from the first epoch was used to calibrate the FIRST baseline's orientation. The results are presented in Table~\ref{table:results}. The error presented in the Table is computed as the standard deviation of the cubes averaged estimates divided by $\sqrt{(N-1)}$, with N the number of cubes. We also need to take into consideration a systematic error on the plate scale. It is proportional to the baseline length divided by the wavelength, and leads to an error of about $1$~mas on the separation estimation. 

\begin{figure}[!h]
	\centering
	\includegraphics[width=0.95\linewidth]{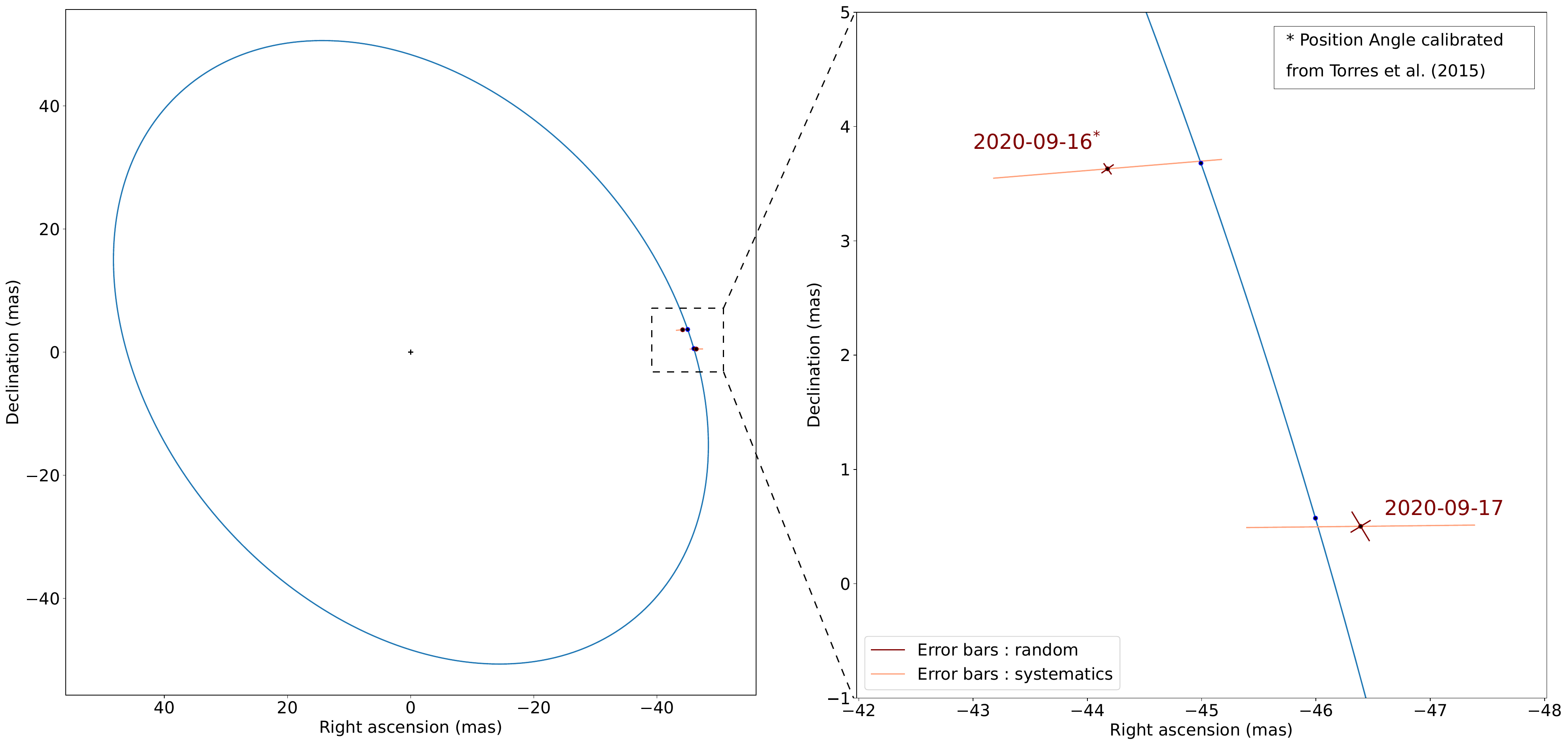}\\
	\caption{Orbit of Hokulei Ab binary component. In blue and brown are respectively the expected and estimated positions of the Ab component on both observation days. Brown colored error bars only take into account the random errors. Salmon colored error bars are a quadratic sum of random and systematic errors.}
	\label{fig:Capella_pos}
\end{figure}

We plot in Fig.~\ref{fig:Capella_pos} the orbit of Hokulei Ab, and show the expected and estimated position of the Ab binary component for both dates of observation. Both errors are presented on the graph. The orientation of the random error is computed from the data covariance presented in Table~\ref{table:results}. As we can see, the estimations compare well with the theoretical positions. If some work remains to reduce the systematics, the PA estimation is accurate down to about $0.1$~degree: the expected PA difference between the two epochs is $4.0$~degrees, and the estimated differential is $3.9$~degrees.

\begin{table}[h!]
		\centering
	\caption{Summary of the results.}
	\label{table:results}
	\begin{tabular}{cccccc}
		\hline \hline
		 Date & Expected     &     Expected      & Estimated    &     Estimated    & Data \\
           & separation   &  position angle   & separation   &  position angle  & covariance     \\
		\hline 
		Sept. 16th 2020  &  $45.1$ mas  &  $274.7$ degrees & $44.3\pm0.1$ mas &  $274.7\pm0.1$ degrees & $-0.02 \text{ mas}^2$ \\ 
		Sept. 17th 2020  &  $46.0$ mas  &  $270.7$ degrees & $46.4\pm0.1$ mas &  $270.6\pm0.2$ degrees & $ 0.32 \text{ mas}^2$ \\
		\hline \hline
		\end{tabular}
\end{table}
{}

\subsection{Interferometric wavefront sensing with FIRST}
\label{iwfs-sec}

Since single aperture interferometers naturally measure relative phase between subpupils, they are the perfect candidates to measure phase discontinuities in the pupil, and help with the pupil fragmentation issue. Pupil fragmentation can result either from spider shadow or segmented primary mirrors. In both cases, differential piston between the fragments lead to degraded high contrast instrument performance, since current Extreme Adaptive Optics sensors cannot properly sense them. We propose here to use FIRST as an interferometric wavefront sensor to help sensing the piston modes between the pupil fragments of the Subaru telescope.

\begin{figure}[!h]
	\centering
	\includegraphics[width=0.8\linewidth]{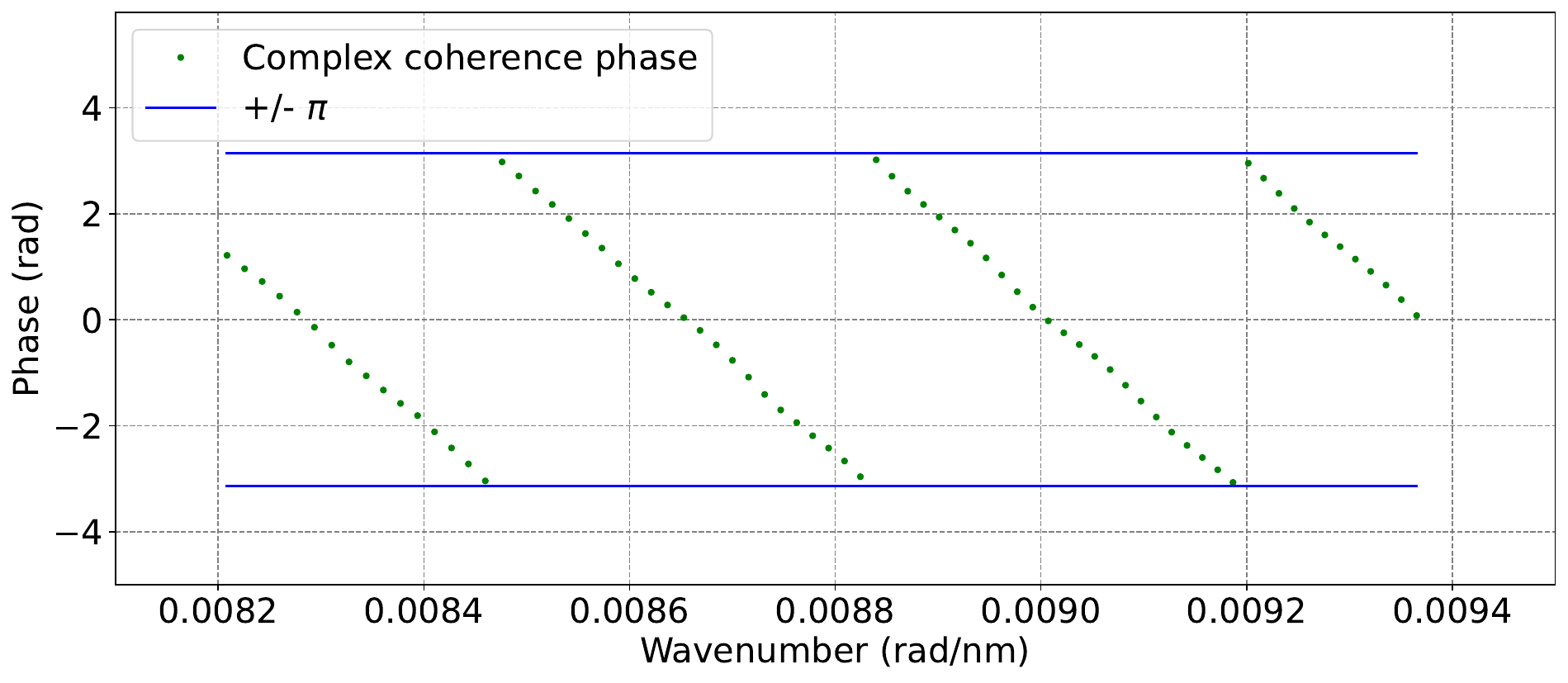}\\
	\caption{Phase of the complex coherence between subpupils linked to segments 37-24 as a function of the wavenumber defined as $2\pi/\lambda$ (green). The phase is wrapped between $\pm\pi$, as materialized by the blue lines.}
	\label{phase_segs-1}
\end{figure}

The complex coherence formed by subpupils $n$ and $n'$ is written as :
\begin{align}
	\mu_{nn'}=|V_{nn'}|\text{e}^{i\psi_{nn'}}A_nA_{n'}\text{e}^{i\Delta\Phi_{nn'}}.\label{eq:mu}
\end{align}
with $|V_{nn'}|$ and $\psi_{nn'}$ the objects visibitily and phase respectively. $A_n$ and $A_{n'}$ are each subpupil intensity and $\Delta\Phi_{nn'}$ is the baseline phase. In the case of a point source, the objects complex visibility modulus is $1$, and its phase is $0$ or $\pi$. Eq.~(\ref{eq:mu}) then becomes :
\begin{align}
    \mu_{nn'}=A_nA_{n'}\text{e}^{i\Delta\Phi_{nn'}}.
\end{align}
The baseline phase is written as: 
\begin{align}
\Delta\Phi_{nn'} = 2\pi\sigma\delta_{nn'} + 2k\pi  \text{ (k $\in$ $\mathbb{N}$)} \label{eq:phase_opd}
\end{align}
with $\sigma$ the wavenumber, in nm\textsuperscript{-1}, and $\delta_{nn'}$ the optical path difference (OPD) between the two apertures, in nm. We acquired data on the SCExAO internal calibration source to test the phase extraction. We recorded frames of 50 ms exposure each without any perturbation. We plot in Fig.~\ref{phase_segs-1} this phase as a function of the wavenumber for the baseline formed by subpupils $37-24$, for one frame (the first frame). We can see that the phase is wrapped, which is expected when the OPD between the subpupils is large. 

{}

$\delta_{nn'}$ is then extracted from the baseline phase, after unwrapping, following :

\begin{equation}
     \delta_{nn'} = \frac{1}{2\pi}\frac{\partial\Delta\Phi_{nn'}}{\partial\sigma}
\end{equation}

\begin{figure}[h!]
    \centering
    \includegraphics[width=0.9\linewidth]{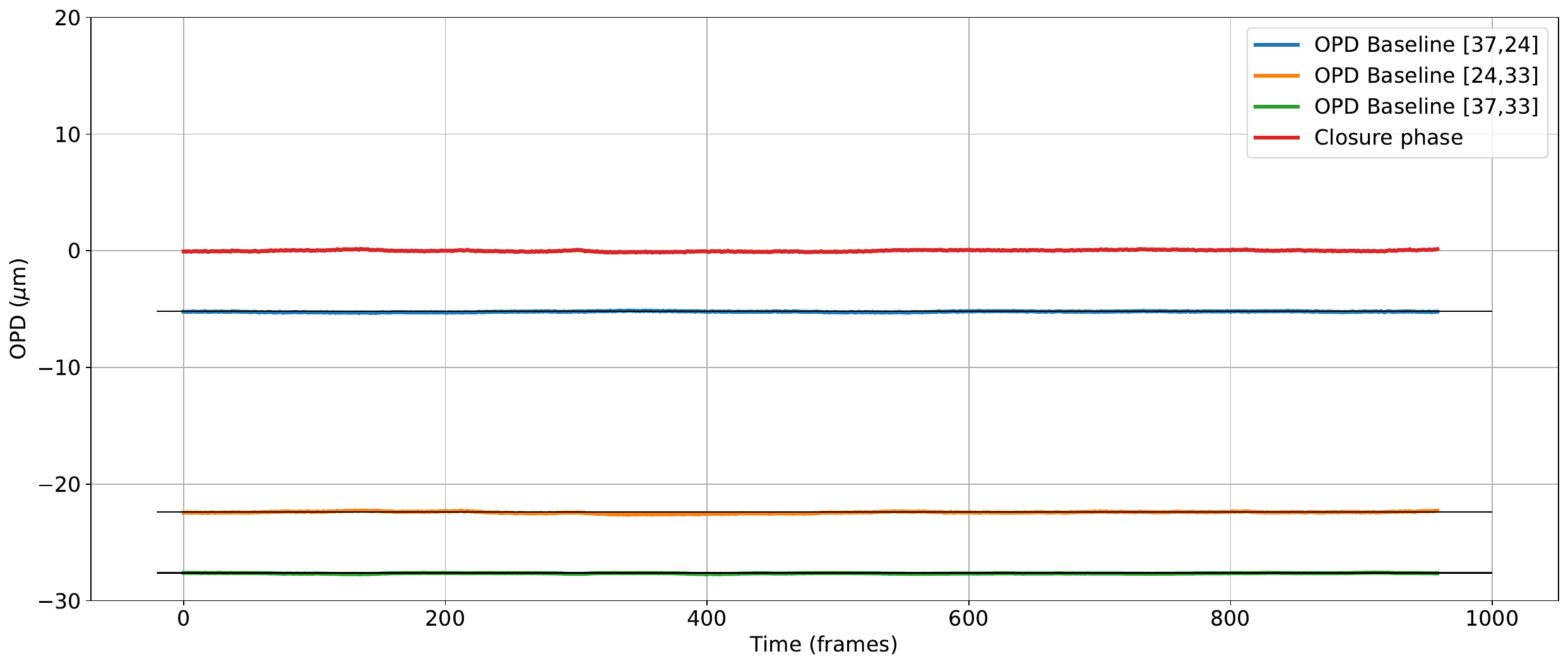}
    \caption{Evolution of the optical path difference for three baselines over time without petaling modes.}
    \label{OPD_segs}
\end{figure}

This operation is done for every baseline, and every frame. Figure~\ref{OPD_segs} shows the evolution of the OPD for three different baselines formed by subpupils $37-24-33$, over the 500 frames. For each baseline, the OPD varies around an average value that is plotted with a solid black line. The variations around the constant average value are caused by turbulence on the SCExAO bench plus thermal/mechanical instabilities. We assume that these variations average to zero over the frames, allowing to state that the OPD average value corresponds to the static path length mismatch between the fibers. The measured standard deviation for the $[37,24]$, $[37,29]$ and $[24,29]$ baselines are respectively $41.1$, $74.3$ and $28.3$~nm~RMS. The Peak-To-Valley variations can reach a couple hundreds of nanometers, which can limit performance in the context of wavefront sensing, and justifies the installation of a metrology source to calibrate the systems instabilities (see Sect. \ref{sec-metrology}). 

{}

\begin{figure}[h!]
    \centering
    \includegraphics[width=0.9\linewidth]{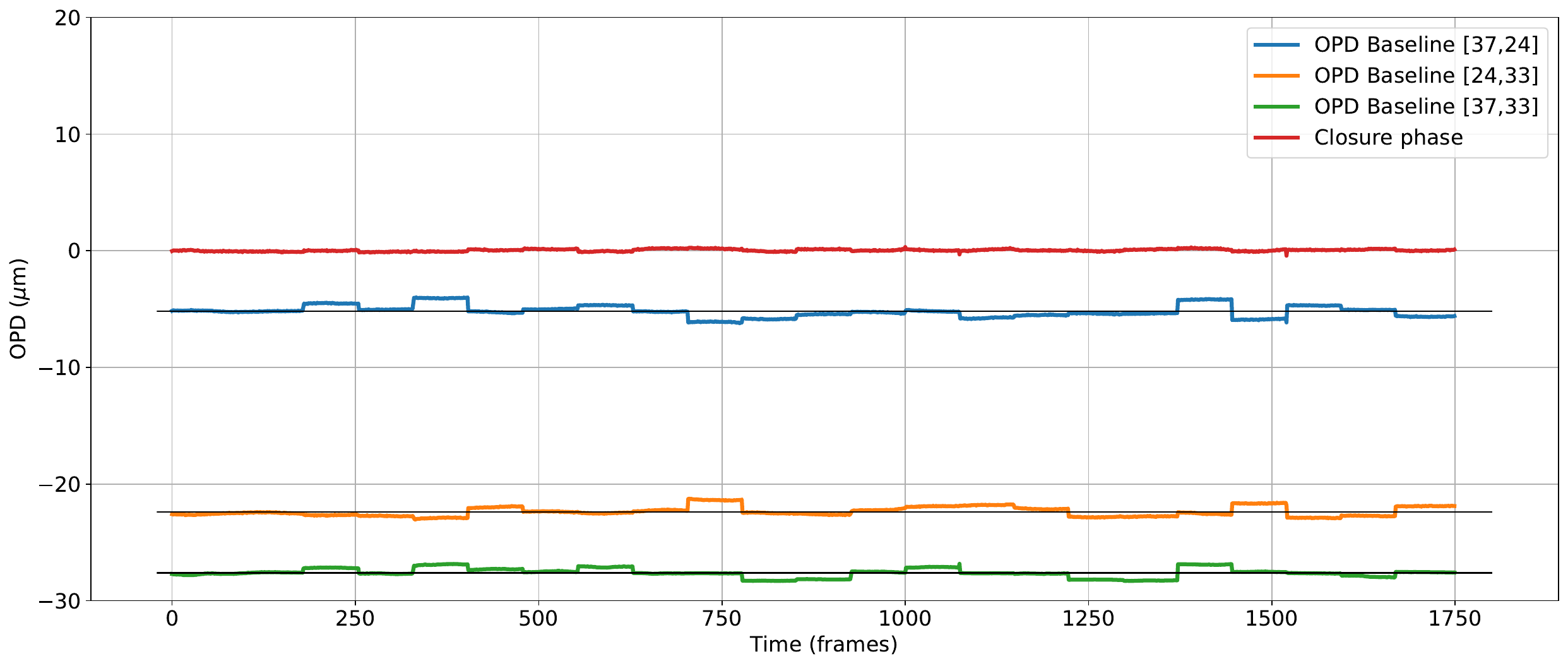}
    \caption{Evolution of the optical path difference for three baselines over time with petaling modes applied on SCExAO DM.}
    \label{fig:first-SuperK-test}
\end{figure}

We performed a second test where we successively applied petaling modes (differential pistons between the pupil fragments) on the SCExAO DM. We show the evolution of the phase as a function of time in Figure~\ref{fig:first-SuperK-test}.

{}

We can see steps in the OPD phase, signature of the petal modes being applied on the SCExAO DM. This work is on-going and a proper phase map reconstruction algorithm is being developed. However this quick tests shows that our instrument is sensitive to differential piston between the pupil fragments.



\section{FIRST upgrades for H$\alpha$ spectro-imaging}

Upgrades aim at increasing the sensitivity and accuracy of the instrument. A first goal is, as stated in the introduction, to identify accreting planets thanks to H$\alpha$ imaging. Even though it is out of our field of view, we take as a reference the AB Aurigae protoplanet candidate whose contrast is $10^{-3}$ and R magnitude 7. We summarize in the following table the current capabilities of FIRST and the goals of our upgrades.

\begin{table}[h]
\centering
\begin{tabular}{|c|c|c|}
	\hline 
                        & Current FIRST & FIRST upgrades \\ \hline
	Number of subpupils & \multicolumn{2}{c|}{9} \\ \hline 
        Number of baselines & \multicolumn{2}{c|}{36} \\ \hline 
        Spectral range & \multicolumn{2}{c|}{600 - 800 nm} \\ \hline
        Spectral resolution & 300 @ 700 nm & \upgrades{\textbf{4000 @ 650 nm}}  \\ \hline 
	  Total Field of View & \multicolumn{2}{c|}{136 mas @ 700 nm} \\ \hline 
        Optimal field of View & \multicolumn{2}{c|}{20 mas @ 700 nm}\\ \hline
        Spatial sensitivity & \multicolumn{2}{c|}{down to 5 mas} \\ \hline 
	Achievable contrast & $2\times 10^{-2}$  & \upgrades{\boldmath{$1\times 10^{-3}$}} \\ \hline
        Magnitude limit & 3 & \upgrades{\textbf{7}} \\
	\hline 
\end{tabular}
\caption{Current and targeted characteristics/performance of FIRST.}
\end{table}

\subsection{Injection upgrades}

\subsubsection{Improving the throughput}
To increase the throughput, we first decrease the number of optical surfaces and non-polarization optics. The combination Polarized Beam-splitter cube + Quarter-wave plate (i.e. 10 crossed interfaces) is upgraded to a linear polarizer + D-mirror (i.e. 3 crossed interfaces). In the future, we plan on completely getting rid of the linear polariser to exploit both polarizations in our data.


\subsubsection{Upgrade of the MEMS}
The MEMS plays a key role in optimizing the injection of the subpupils light into the single-mode fibers. We have upgraded the MEMS device from an Iris AO to a Boston Micromachine while keeping the same number of segments (37). The difference between the two lies in the size of the device. This forced us to change all the optics of the injection, to adapt our setup to the new size of the device. The following Table~3 presents the optical features of the old setup compared to the new setup.

\begin{table}[h]
\centering
\begin{tabular}{|c|c|c|}
	\hline 
                                      & Old injection & New injection     \\ \hline \hline 
	NF-number (from SCExAO)       & \multicolumn{2}{c|}{f/27.3}           \\ \hline 
        Collimation lens focal length & 120 mm        & 125 mm   \\ \hline 
        Beam size on collimation lens & 4.24 mm       & 4.58 mm  \\ \hline
        MEMS                          &    Iris AO    & Boston Micromachines \\ \hline 
        Number of Hex-segments        & \multicolumn{2}{c|}{37}           \\ \hline 
        MEMS - full aperture size     & 4.24 mm       &  4.55 mm \\ \hline 
        MEMS - inscribed circle in each segment & 606.2 $\mu$m &  649.5 $\mu$m\\ \hline 
        First telescope lens         & 85 mm         & 88.9 mm \\ \hline 
        Second telescope lens        & 35 mm         & 35 mm  \\ \hline
        Pupil size on MLA             & 1.75 mm       & 1.79 mm \\ \hline 
        Subpupil size on each $\mu$-lens & 250 $\mu$m & 255.7 $\mu$m \\ \hline
        $\mu$-lens diameter           & \multicolumn{2}{c|}{250 $\mu$m }   \\ \hline
        $\mu$-lens focal length       & \multicolumn{2}{c|}{1 mm }\\ \hline 
        Fiber bundle pitch            & \multicolumn{2}{c|}{250 $\mu$m }   \\ \hline
\end{tabular}
\label{table_first}
\caption{FIRST injection module optical elements before and after the MEMS upgrade.}
\end{table}

\subsubsection{Metrology source}
\label{sec-metrology}

\begin{figure}[!h]
    \centering
    \includegraphics[width=0.7\linewidth]{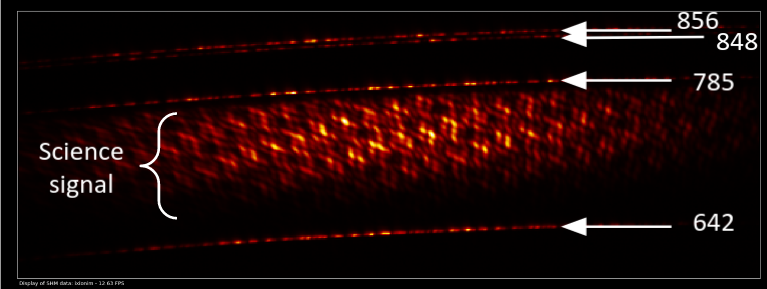}
    \caption{Raw image of the FIRST spectrograph where we image both the SCExAO calibration broadband laser (SuperK) and the FIRST internal metrology.}
    \label{fig:first-metro}
\end{figure}

As mentioned in Section~\ref{iwfs-sec}, the instabilities of the fibers can/will be a problem when it comes to accurately measuring the phase between the subpupils. In order to disentangle the phase component coming from the instrument from the contribution of turbulent residuals or low wind effect, we installed a metrology system comprising four lasers at 642~nm, 785~nm, 848~nm and 856~nm. These metrology sources are injected in parallel to the science light: incoming beam from SCExAO is reflected towards FIRST thanks to a R90/T10 filter (FIRST pickoff in Figure \ref{fig:first-scexao-scheme}). We use the filter in transmission to simultaneously inject light coming from the metrology system.
We provide in Figure~\ref{fig:first-metro} an example of a raw image obtained when injecting the SCExAO SuperK internal source in parallel to the metrology sources.

{}

\begin{figure}[!h]
    \centering
    \includegraphics[width=0.8\linewidth]{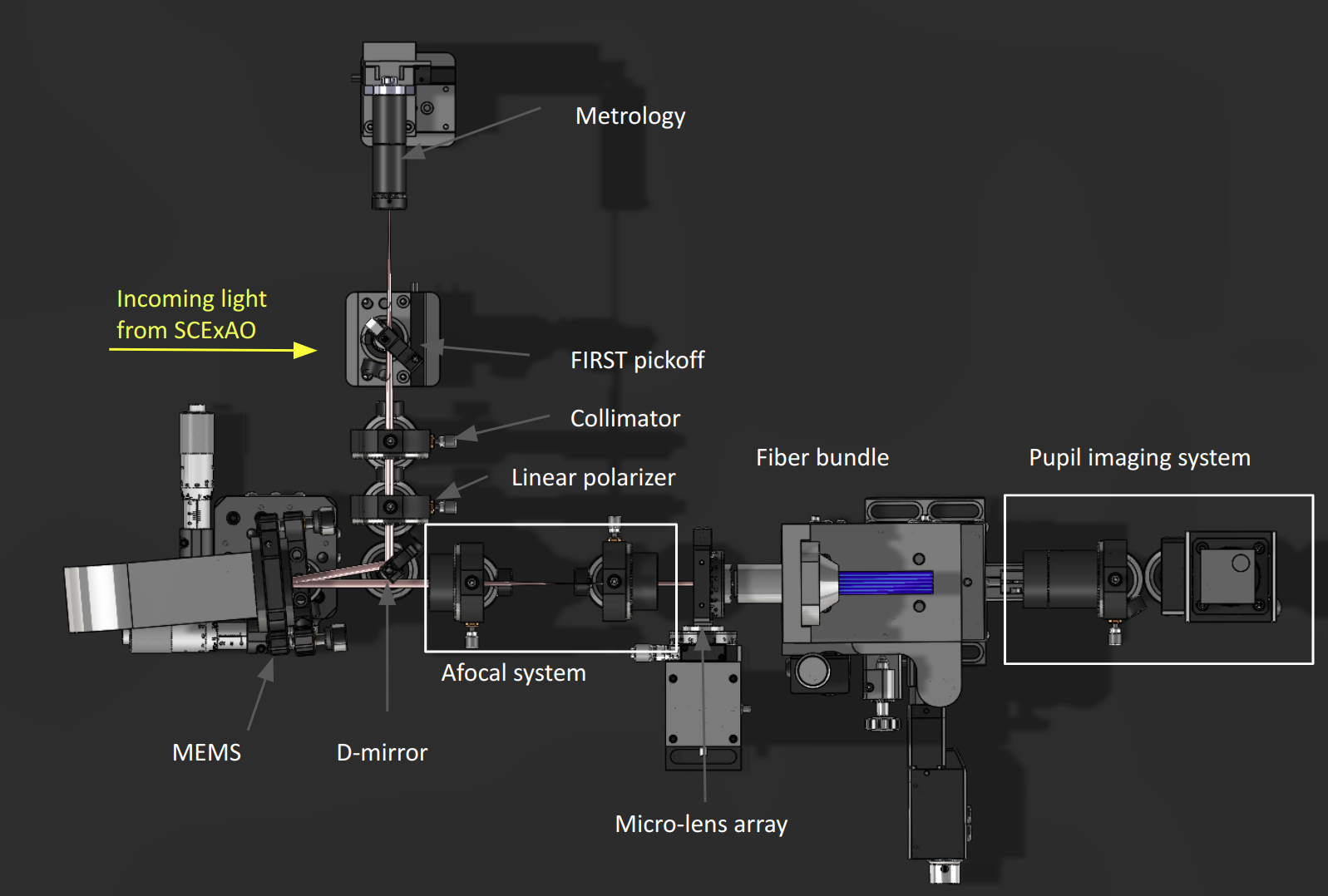}
    \caption{Upgraded FIRST injection module with a new MEMS, new optical setup including a D-mirror and a metrology source system.}
    \label{fig:first-first-newinjection}
\end{figure}

The upgraded injection module CAD design is displayed in Figure~\ref{fig:first-first-newinjection}.

{}

\subsection{Recombination upgrades}

\subsubsection{Integrated chip}

\begin{figure}[!h]
	\includegraphics[width=\linewidth]{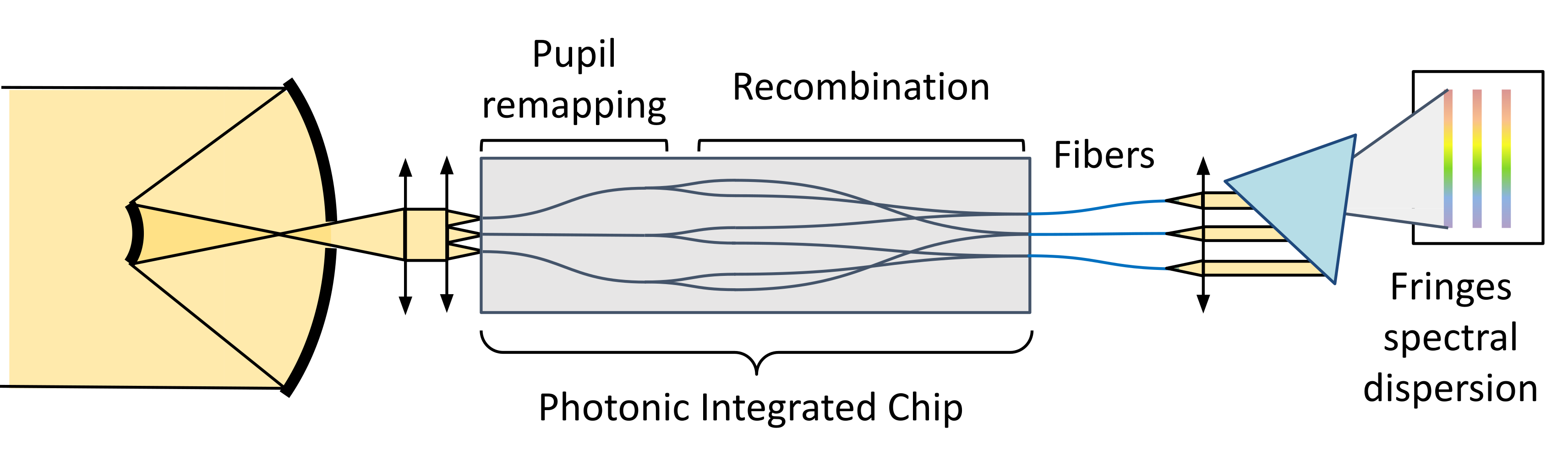}\\
	\caption{FIRST Photonic Integrated Chip approach principle.}
	\label{fig-FIRST-PIC}
\end{figure}

For the subpupil recombination to be more stable, accurate and sensitive, we aim at deploying a Photonic Integrated Chip (PIC). This method has been successfully demonstrated on-sky for exoplanet detection using VLT/GRAVITY~\cite{lacour2019first}. It has also been successfully demonstrated for nulling interferometry with the Subaru/GLINT instrument~\cite{martinod2021scalable}. The single mode fibers are connected to the PIC where single-mode waveguides are engraved. The waveguides split and recombine the subpupils pair-wise. A simplified scheme of such design is provided in Fig.~\ref{fig-FIRST-PIC}. Because our fibers have a non-negligible optical path difference, we added optical delay lines upstream from the PIC, in order to phase the subpupil beams.

{}

While this kind of device is relatively well manufactured and exploited in the NIR, their manufacturing in the visible remains difficult. Work towards a better understanding of these devices behavior in the Visible has been initiated with TEEM Photonics over the last few years~\cite{10.1117/12.2312841,10.1117/12.2630039}. Multiple PIC designs~\cite{lallement2023photonic} have recently been manufactured, and are being tested on a dedicated testbed in Paris Observatory~\cite{barjot2021}. In the next two paragraphs, we present characterization results of the latest PIC prototype, hereafter referred to as 5TX PIC, a 5 sub-apertures beam combiner using directional couplers. 

\paragraph{Improved cross-talk performance and measurement of the polarization cross-talk:} to estimate the cross-talk occurring between waveguides (called cross-talk) and the cross-talk occurring between polarizations (called polarization cross-talk), we take a set of frames when injecting light into a single input, this set is hereafter called a "flat". We take one flat per input when injecting V-polarized light into the PIC. V-polarized (or H-polarized) light is obtained using a linear polarizer located upstream of the fibers injecting light into the PIC. If the physical orientation of the linear polarizer is vertical (resp. horizontal) with respect to the bench surface, the injected light is called V-polarized (resp. H-polarized). For the combination scheme of a 5TX, ideally we expect to see all the light equally distributed in 8 outputs and no light in the 32 other outputs. K.Barjot~\cite{barjot2021} previously measured a mean cross-talk of 1$\%$ and of 10$\%$ in the worst case. The goal is to achieve cross-talk and polarization cross-talk of less than $1\%$ in the worst case. \\

\begin{figure}[!h]
\centering
	\includegraphics[width=0.7\linewidth]{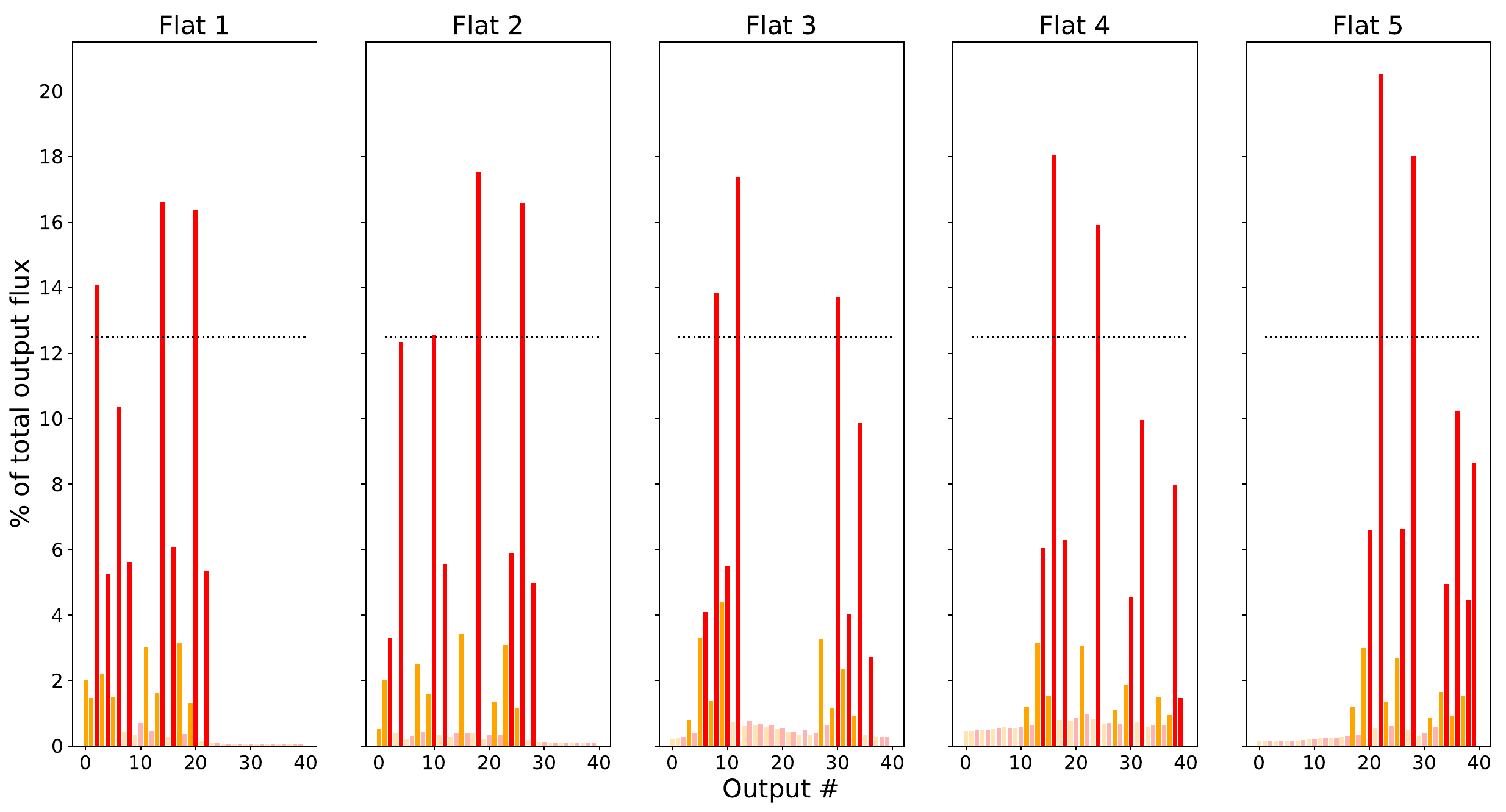}\\
	\caption{Percentage of the total output flux measured in each output of the new 5TX PIC prototype. Dark red bars correspond to the V-polarized part of the light coming from the selected outputs, where we expect 12.5$\%$ of the total output flux. Dark orange bars correspond to the H-polarized part of the light coming from the selected outputs, where we expect 0$\%$ of the total output flux due to the V-polarized light injection. Light red (resp. orange) bars represent the amount of V-polarized (resp. H-polarized) light coming from the unselected outputs, where we also expect 0$\%$ of the total output flux.}
	\label{fig:FIRST-Crosstalk}
\end{figure}

Figure.~\ref{fig:FIRST-Crosstalk} presents the percentage of the total output flux measured in each output of the 5TX PIC prototype. It shows that for the selected outputs and polarization (dark red bars), the percentage of the total output flux has an average value of 9.50$\%$ instead of 12.5$\%$. This discrepancy can be explained by losses due to propagation, cross-talk and polarization cross-talk. Cross-talk corresponds to light leakage in outputs where light is not expected (light red or orange bars). We measure an average value of 0.36$\%$ and a worst case value of 0.98$\%$ across all flats. Polarization cross-talk corresponds to the amount of H-polarized light coming from the selected outputs (dark orange bars) which has an average value of 1.93$\%$ over all flats. The worst value is 4.41$\%$. We reduced the mean cross-talk from 1$\%$ ($10\%$ at worst) to 0.36$\%$ (0.98$\%$ at worst) which is within specification. We also measured a mean polarization cross-talk of 1.93$\%$ (4.41$\%$ worst case) which needs further optimization to get below 1$\%$.

{}

\subsubsection{Spectrograph}
\label{sec-spectro}
\begin{figure}[!h]
    \centering
    \includegraphics[width=\linewidth]{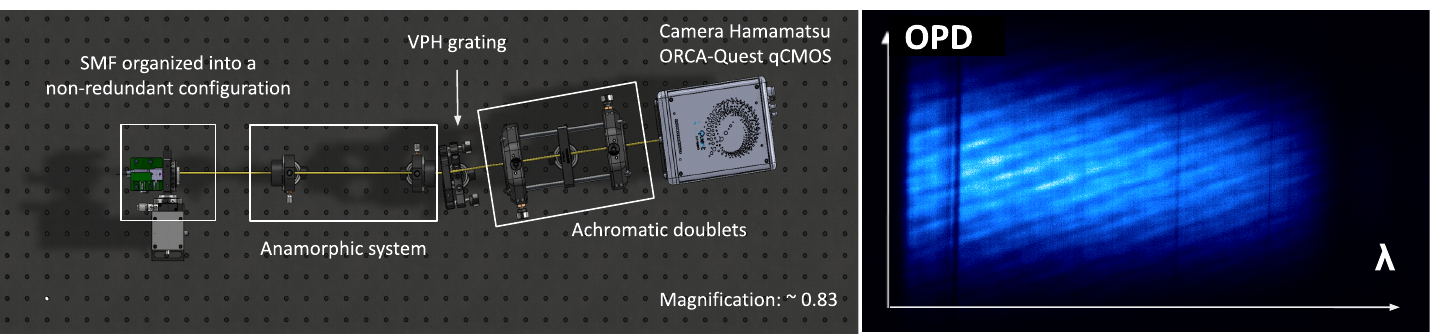}
    \caption{New R-4000 spectrograph integrated in FIRST. Left: CAD rendering of the FIRST anamorphic system followed by the spectrograph. Right: On-sky testing of the new spectrograph. Acquisition on Hōkūle'a [\textit{Arcturus}]}
    \label{fig:first-r4000onsky}
\end{figure}

The new spectrograph was designed to be fed by single-mode fibers positioned into a V-groove with a pitch of 127~$\mu$m. A x2 apochromatic microscope objective collimates the fiber output beams before they get dispersed by a volume holographic grating manufactured by Wasatch Photonics. Two achromatic doublets then image the fiber outputs on a Hamamatsu ORCA-Quest camera, reaching a resolution of about 4000. 
Although the spectrograph was originally designed to image the outputs of the PIC previously mentioned, it was also deployed on the current Subaru/FIRST setup in order to enhance the resolution capabilities of the current instrument, in parallel to the PICs developments led in Paris Observatory. Figure~\ref{fig:first-r4000onsky}-left shows the CAD drawing of the spectrograph integrated after the anamorphic system of FIRST. 
Figure~\ref{fig:first-r4000onsky}-right shows an image acquired on-sky on Hōkūle'a [Arcturus], with the new spectrograph.

{}

\subsection{The Photonic Lantern approach}

\subsubsection{The Photonic Lantern principle}
\begin{figure}[!h]
\centering
	\includegraphics[width=\linewidth]{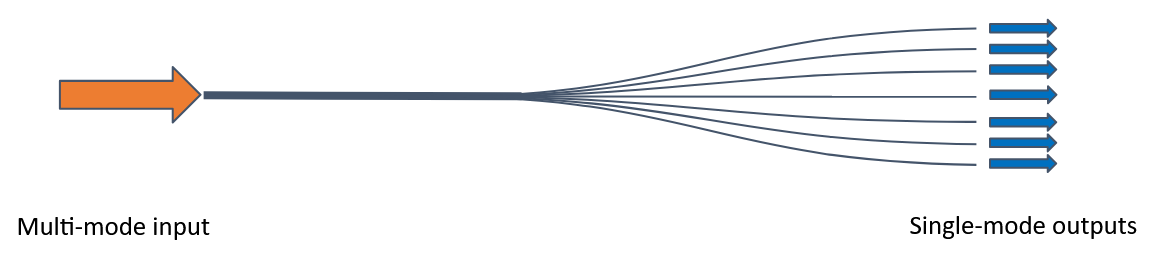}\\
	\caption{Principle of the pigtail photonic lantern approach: light is injected into a multi-mode input. The multi-mode guide slowly becomes several single-mode waveguides at the output.}
	\label{fig-PL-principle}
\end{figure}

Even though our various photonic upgrades might provide more stable interferometric signal, there is still a limitation in terms of throughput. One solution that is investigated in parallel is the use of a photonic lantern (PL)~\cite{leon2010photonic}. PLs are optical fiber devices where the input is a multi-mode (MM) waveguide and the output is an array of single mode (SM) waveguides (see Figure~\ref{fig-PL-principle}). The adiabatic transition between the two modes (MM to SM) is very efficient with a throughput greater than 90\%~\cite{birks2015photonic}.

{}

\subsubsection{The Photonic Lantern as an input of a PIC}
\begin{figure}[!h]
\centering
	\includegraphics[width=\linewidth]{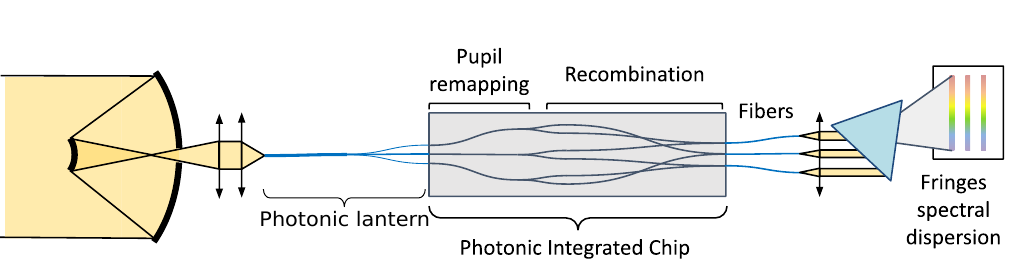}\\
	\caption{Instrument design with a photonic lantern feeding a photonic integrated chip.}
	\label{FIRST-PL-PIC}
\end{figure}

In the context of FIRST, the ultimate idea would be to feed a PIC with the PL outputs. A recent study from Y. J. Kim et al.~\cite{YJKim2023} shows that coherent imaging is possible by recombining the SMF outputs of a PIC. The study demonstrates through simulations that the interferometric observables obtained from the PL SM outputs recombination behave similarly to visibilities obtained with "classical" interferometry. Such instrument design is showed in Figure~\ref{FIRST-PL-PIC}.

{}

Advantages here are twofold. First, the light from the entire pupil can be coupled into the MM waveguide, significantly boosting the sensitivity, compared to FIRST which can only couple light from a fraction of the pupil. Second, coupling light into a MM core is much easier than in a SM core - especially in the Visible wavelengths where the alignment requirements are high, and high Strehl ratios difficult to achieve. 




\subsubsection{Photonic Lantern testings on SCExAO}

\begin{figure}[!h]
\centering
	\includegraphics[width=\linewidth]{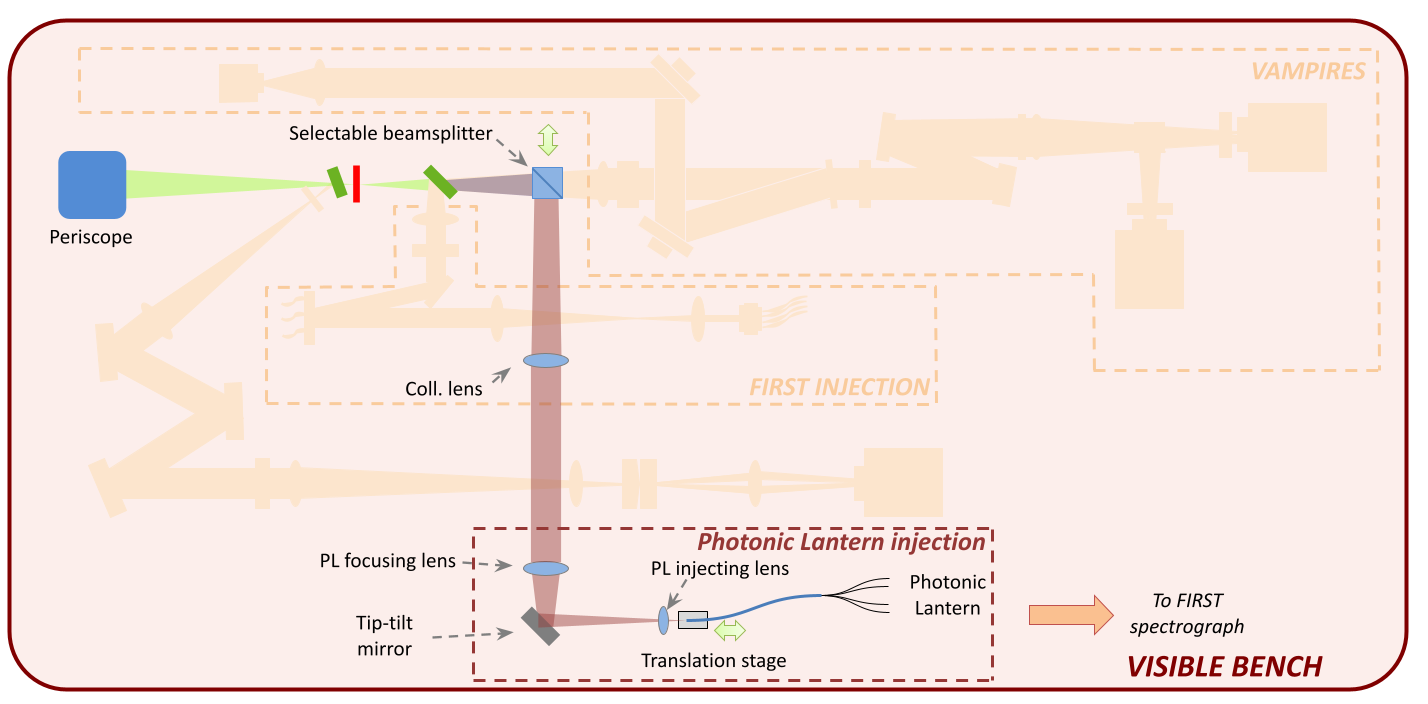}\\
	\caption{Photonic Lantern injection module on the SCExAO Visible bench.}
	\label{fig-PL-}
\end{figure}

Prior to using a PL as an input of a PIC, we have tested the PL device alone. To do this, we built an injection module on the Visible bench of SCExAO, in parallel to FIRST. Figure~\ref{fig-PL-} shows the optical layout of the injection module on the SCExAO Visible bench.

{}

\begin{figure}[!h]
\centering
	\includegraphics[width=0.6\linewidth]{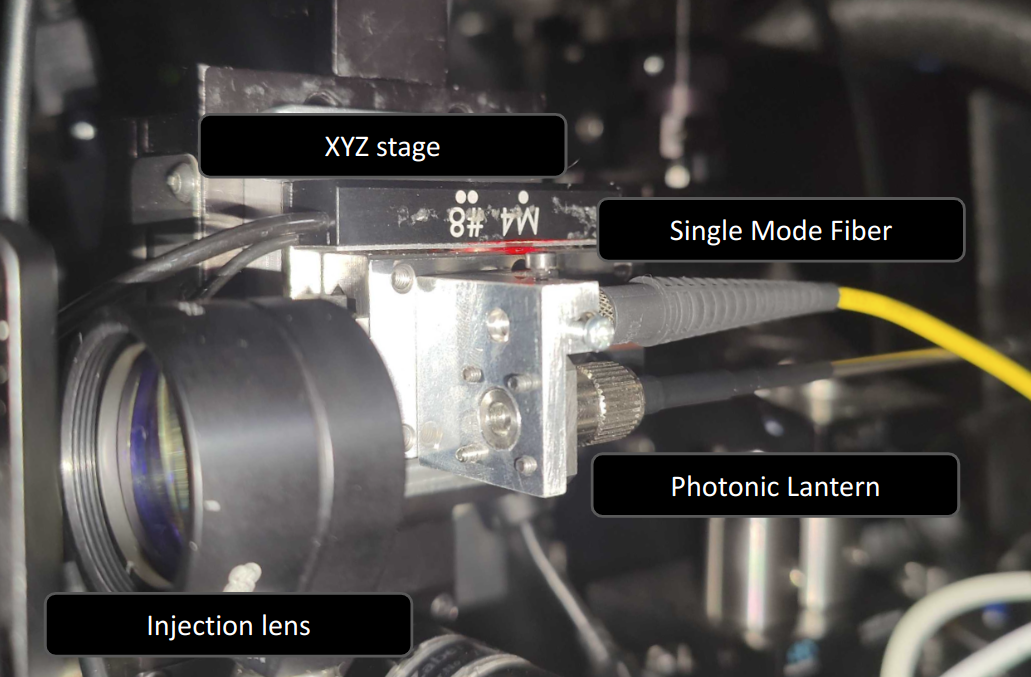}\\
	\caption{Injection module.}
	\label{fig-PL-pic}
\end{figure}

A beamsplitter cube is used to send the light toward a collimation lens plus a system of two lenses that inject light into the PL or in an SMF (see Figure~\ref{fig-PL-pic}). The SMF is used to check the quality of the focal plan after the injection lens, and for comparison with the PL.

{}

\begin{figure}[!h]
\centering
	\includegraphics[width=0.9\linewidth]{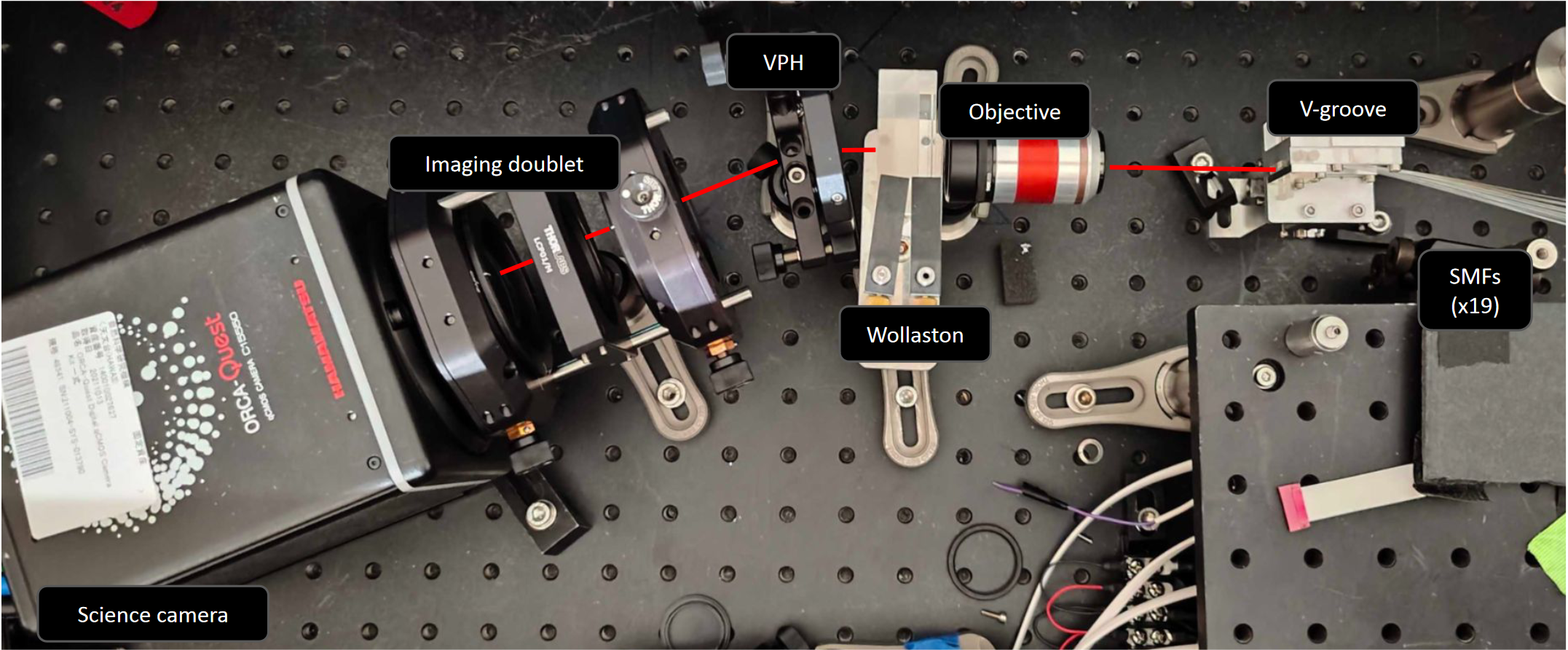}\\
	\caption{Spectrograph imaging the 19 SMF outputs of the photonic lantern.}
	\label{fig-PL-pic-spectro}
\end{figure}

Both fibers run to the FIRST recombination bench described in Section~\ref{sec-firstrecomb}. The SMF output is imaged on the FIRST photometric monitoring arm. The PL SMF outputs are spliced into a V-groove with a pitch of $127~\mu m$. The V-groove outputs are imaged on a spectrograph which is a replica of the FIRST upgraded spectrograph presented in Section~\ref{sec-spectro}. Figure~\ref{fig-PL-pic-spectro} presents the spectrograph imaging the 19~SMF outputs of the PL. In this setup, we add a Wollaston, used to split both polarizations of the light. Therefore, 38~spectra are recorded: 2 polarizations per output.

{}

\subsubsection{On-sky testing}

\begin{figure}[!h]
\centering
	\includegraphics[width=\linewidth]{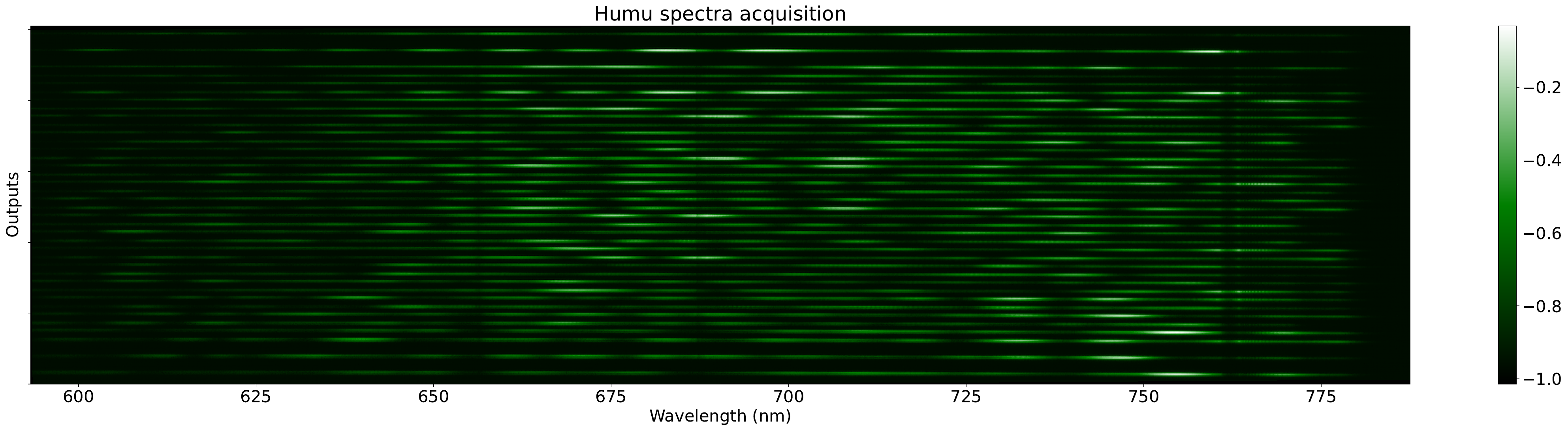}\\
	\caption{Spectra of Humu acquired with the Photonic Lantern. The horizontal axis is the wavelength dispersion, while the 38 output traces, corresponding to the two polarizations from each of the PL 19 SMF outputs, spread along the vertical axis.}
	\label{fig:Humu-im}
\end{figure}

\begin{figure}[!h]
\centering
	\includegraphics[width=\linewidth]{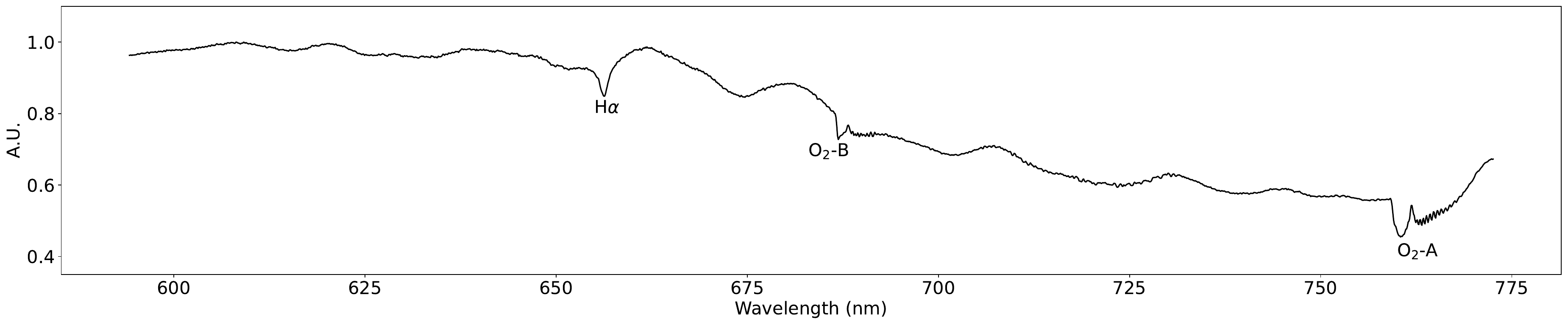}\\
	\caption{Spectrum of Humu acquired with the Photonic Lantern after co-adding all output spectra.}
	\label{fig:Humu-spectrum}
\end{figure}

We tested the photonic lantern on-sky during an Engineering run on July 30th 2023. Figure~\ref{fig:Humu-im} shows the dark-subtracted average of 120,000 frames obtained on Humu [Altair] with a framerate of 200~Hz. Polarization splitting is vertical on the image, and spectral dispersion is horizontal. We can identify from this image several absorption lines. We compute the spectrum of Humu by co-adding all the output signals - after individual extraction of the traces. Each output response is calibrated using a halogen lamp illuminating the PL input uniformly, allowing to record flat-field data. Before co-addition of the traces, they were all divided by their individual flat-field. The resulting spectrum of the calibrated and co-added data is shown in Figure~\ref{fig:Humu-spectrum}.

{}
{}

The spectrum of Humu shows three different absorption lines: Oxygen bands A and B from the atmosphere (around 687~nm and 761~nm), and H$\alpha$ (at about 656~nm). This result shows that our setup can be used for high-throughput H$\alpha$ spectro-imaging. This demonstration constitutes the first on-sky acquisition of a spectrum using a photonic lantern fed by an Extreme Adaptive Optics system in the Visible (Vievard et al., in prep). More tests and characterization of the device will be done in the near future.

\section{Conclusion}

We presented the FIRST spectro-interferometer operating behind an ExAO system in the Visible (600-800 nm). FIRST samples and recombines several parts of a telescope pupil, with spectral dispersion capabilities (R$\sim300$ at 700 nm). 
FIRST is routinely tested on-sky. Performance evaluation on unresolved source observation showed that FIRST should be sensitive to spatial structures down to a quarter of the telescope spatial resolution. We also assessed that the instrument can enable detection with a contrast down to 0.02 at $\lambda/D$ separation. The observation of Hokulei demonstrated FIRST's capability to detect and track the position of the binary companion along its orbit using observations about 24~hours apart. The best random error on the separation and PA were respectively $0.1$~mas and $0.1^\circ$. Systematic errors on the separation remain due to poor calibration of the subpupil projected diameter. More tests on well-known binaries will help to calibrate these systematic errors.  
The design of the instrument makes it also sensitive to discontinuous aberrations in the pupil, like differential pistons due to petalling/Low Wind Effect. We showed that the differential piston information is contained in the complex coherence of the baseline, and needs to be extracted properly. 

We also presented several on-going upgrades on the instrument aiming at increasing its sensitivity and stability, with the goal of performing H$\alpha$ spectro-imaging. Injection was upgraded to reduce optical surfaces, and we added a metrology source, injected in parallel to the science beam, to disentangle the phase coming from the fiber instabilities and the turbulent wavefront residuals. On the recombination part, we upgraded the spectrograph to increase the resolution by a factor of 10. We are also exploring the use of Photonic Integrated Chips for the recombination. Several devices are tested in the lab at the Paris Observatory. The goal is to increase their throughput and cross-talk performance. Finally, we introduced a new photonic solution we started exploring in parallel, that is the Photonic Lantern. Thanks to its MM input and SM outputs, it can be used as a high throughput high resolution instrument. A visible PL was installed on SCExAO and is feeding a replica of the FIRST R$\sim3000$ spectrograph. On-sky testings are promising and constitute the first experiment of feeding a photonic lantern device with ExAO-corrected light in the visible. 

The success of demonstrating such instrument on SCExAO is an important stepping stone for future photonic instrumentation on extremely large telescopes. Their large collecting area and small diffraction limit (about 4~mas in the optical wavelengths), coupled with spectral dispersion capabilities, would offer unique high contrast and high resolution capabilities with high sensitivity for Earth-like exoplanet research and characterization.

\acknowledgments 
The development of FIRST was supported by Centre National de la Recherche Scientifique CNRS (Grant ERC LITHIUM - STG - 639248) and by the French National Research Agency (ANR-21-CE31-0005). The development of SCExAO was supported by the Japan Society for the Promotion of Science (Grant-in-Aid for Research \#23340051, \#26220704, \#23103002, \#19H00703 \& \#19H00695), the Astrobiology Center of the National Institutes of Natural Sciences, Japan, the Mt Cuba Foundation and the director's contingency fund at Subaru Telescope. GLINT work was supported by the Australian Research Council Discovery Project DP180103413. 
Critical fabrication for GLINT was performed in part at the OptoFab node of the Australian National Fabrication Facility utilising Commonwealth as well as NSW state government funding. 
S. Gross acknowledges funding through a Macquarie University Research Fellowship (9201300682) and the Australian Research Council Discovery Program (DE160100714). 
N. Cvetojevic acknowledges funding from the European Research Council (ERC) under the European Union’s Horizon 2020 research and innovation program (grant agreement CoG - 683029). M.L acknowledges support from the doctoral school Astronomy and
Astrophysics of Ile de France (ED 127) and from the French National Research Agency (ANR-21-CE31-0005).
The authors wish to recognize and acknowledge the very significant cultural role and reverence that the summit of Maunakea has always had within the indigenous Hawaiian community. We are most fortunate to have the opportunity to conduct observations from this mountain.

\bibliography{main} 
\bibliographystyle{spiebib} 

\end{document}